\documentclass[aps,preprint]{revtex4}%
\usepackage{amsfonts}
\usepackage{amsmath}
\usepackage{amssymb}
\usepackage{graphicx}%
\begin{document}
\title{One-step generation of high-quality squeezed and EPR states in cavity QED. }
\author{C. J. Villas-B\^{o}as and M. H. Y. Moussa}
\affiliation{Departamento de F\'{\i}sica, Universidade Federal de S\~{a}o Carlos, P.O. Box
676, S\~{a}o Carlos\textit{, 13565-905, S\~{a}o Paulo, }Brazil.}

\begin{abstract}
We show how to generate bilinear (quadratic) Hamiltonians in cavity quantum
electrodynamics (QED) through the interaction of a single driven three-level
atom with two (one) cavity modes. With this scheme it is possible to generate
one-mode mesoscopic squeezed superpositions, two-mode entanglements, and
two-mode squeezed vacuum states (such the original EPR state), without the
need for Ramsey zones and external parametric amplification. The degree of
squeezing achieved is up to 99\% with currently feasible experimental
parameters and the errors due to dissipative mechanisms become practically negligible.

\end{abstract}

\maketitle

The preparation of Einstein-Podolsky-Rosen (EPR) entanglement \cite{EPR} has
been a challenge for theoretical and experimental physics since the
introduction of the Bell inequalities \cite{Bell} to test for nonlocal
correlations. Over the last few decades a plethora of experimental
confirmations of nonlocal correlation has contributed to widening the
perspectives for applications of this fundamental phenomenon, from quantum
teleportation \cite{Telep} to computation \cite{Comp}. Equally impressive are
the applications proposed for squeezed states, ranging from fundamental
physics to technology. Possible ways of measuring gravitational waves through
squeezed fields \cite{Caves} and of deepening our understanding of the
properties of radiation \cite{Stoler} and its interaction with matter
\cite{Milburn} have been pursued alongside the preparation of such
nonclassical states.

The preparation of squeezed light, supplied by nonlinear optical media as
running waves \cite{98} and standing squeezed fields in high-$Q$ cavities and
ion traps, generated through atom-field interaction \cite{MZ}, has already
been investigated. The issue of squeezing an arbitrary previously-prepared
cavity-field state $\left|  \Psi\right\rangle $ had not been addressed until
Refs. \cite{Celso1,Roberto1}. Working with Rydberg atoms in the microwave
regime, in this letter,we present a feasible enhanced scheme to engineer
bilinear Hamiltonians in cavity QED. As in Refs. \cite{Celso1,Roberto1}, this
is accomplished through the interactions of a single three-level atom
simultaneously with a classical driving field and a two-mode cavity. However,
the new scheme exhibits a crucial difference from the two previous ones
\cite{Celso1,Roberto1}, both of which operate only in the weak-amplification
regime, owing to the adiabatic approximation required. Here, both the
parametric up- and down-conversions (PUC and PDC) are also derived for the
strong-amplification regime, where the strength of the bilinear and quadratic
interactions between the cavity modes are considerably increased. Hence, a
smaller atom-field interaction time is required for state engineering purposes
and, consequently, the atom-field dissipative mechanisms become negligible.
These interactions are used to generate superpositions of highly-squeezed
states, two-mode squeezed vacuum states (such as the EPR state), the even and
odd EPR states, and entanglements of coherent states. Such states, generated
without the need for Ramsey zones and external parametric amplification, can
be employed either for fundamental tests of quantum mechanics or to manipulate
quantum information.

\textit{PDC}. Consider the atomic levels in the ladder configuration, as shown
in Fig. 1a. The ground $\left|  g\right\rangle $ and excited $\left|
e\right\rangle $ states are coupled through an intermediate level $\left|
i\right\rangle $. The cavity modes $\omega_{a}$ and $\omega_{b}$ ( $\omega
_{a}+\omega_{b}=2\omega_{0}$) are tuned to the vicinity of the dipole-allowed
transitions $\left|  g\right\rangle $ $\leftrightarrow$ $\left|
i\right\rangle $ and $\left|  e\right\rangle $ $\leftrightarrow$ $\left|
i\right\rangle $ with coupling constants $\lambda_{a}$ and $\lambda_{b}$,
respectively, and detuning $\Delta$ $=\omega_{0}+\omega_{i}-\omega
_{a}=-\left(  \omega_{0}-\omega_{i}-\omega_{b}\right)  $. The desired
interaction between the modes $\omega_{a}$ and $\omega_{b}$ is accomplished by
driving out of resonance the dipole-forbidden atomic transition $\left|
g\right\rangle $ $\leftrightarrow$ $\left|  e\right\rangle $ \cite{Forbidden}
with a classical field of frequency $\omega=2(\omega_{0}-\delta)$ and coupling
constant $\Omega$.{\Large \ }Within the rotating-wave approximation, the
Hamiltonian is given by $H=H_{0}+V(t)$, where (with $\hbar=1$)
\begin{subequations}
\begin{align}
H_{0}  &  =\omega_{a}a^{\dagger}a+\omega_{b}b^{\dagger}b+\omega_{0}\left(
\sigma_{ee}-\sigma_{gg}\right)  +\omega_{i}\sigma_{ii},\label{Eq1a}\\
V(t)  &  =\left(  \lambda_{a}a\sigma_{ig}+\lambda_{b}b\sigma_{ei}%
+\Omega\operatorname*{e}\nolimits^{-i\omega t}\sigma_{eg}+\mathrm{h{.c.}%
}\right)  {.} \label{Eq1b}%
\end{align}
with $a^{\dagger}$ ($a$) and $b^{\dagger}$ ($b$) standing for the creation
(annihilation) operators of the quantized cavity modes, while $\sigma
_{kl}\equiv\left|  k\right\rangle \left\langle l\right|  $ ($k$, $l$ being the
atomic states). In a frame rotating with the driving-field frequency, obtained
through the unitary operator $U=\exp\left[  -i\omega t\left(  a^{\dagger
}a+b^{\dagger}b+\sigma_{ee}-\sigma_{gg}\right)  /2\right]  $, the transformed
Hamiltonian $\widetilde{H}=\widetilde{H}_{0}+\widetilde{V}$ is given by
$\widetilde{H}_{0}=\delta_{a}a^{\dagger}a+\delta_{b}b^{\dagger}b+\delta
(\sigma_{ee}-\sigma_{gg})+\omega_{i}\sigma_{ii}+\left(  \Omega\sigma
_{eg}+\mathrm{h{.c.}}\right)  $ and $\widetilde{V}=\left(  \lambda_{a}%
a\sigma_{ig}+\lambda_{b}b\sigma_{ei}+\mathrm{h{.c.}}\right)  $, where
$\delta_{\ell}=\omega_{\ell}-\omega/2$ ($\ell=a,b$). Assuming that $\delta
\ll\left|  \Omega\right|  $ and defining a new basis for the atomic states
$\left\{  \left|  i\right\rangle ,\left|  \pm\right\rangle =\left(
\pm\operatorname*{e}\nolimits^{i\varphi/2}\left|  g\right\rangle
+\operatorname*{e}\nolimits^{-i\varphi/2}\left|  e\right\rangle \right)
/\sqrt{2}\right\}  $\cite{Solano}, composed of eigenstates of the free atomic
Hamiltonian, we obtain in the interaction picture
\end{subequations}
\begin{align}
\mathcal{H}(t)  &  =\lambda_{a}\operatorname*{e}\nolimits^{i\varphi/2}a\left(
\operatorname*{e}\nolimits^{i(\Delta-\left|  \Omega\right|  -\delta)t}%
\sigma_{i+}+\operatorname*{e}\nolimits^{i(\Delta+\left|  \Omega\right|
-\delta)t}\sigma_{i-}\right)  /\sqrt{2}\nonumber\\
&  +\lambda_{b}\operatorname*{e}\nolimits^{i\varphi/2}b\left(
\operatorname*{e}\nolimits^{-i(\Delta-\left|  \Omega\right|  +\delta)t}%
\sigma_{+i}+\operatorname*{e}\nolimits^{-i(\Delta+\left|  \Omega\right|
+\delta)t}\sigma_{-i}\right)  /\sqrt{2}+\mathrm{h.c.} \label{Eq2}%
\end{align}
where $\Omega=\left|  \Omega\right|  \operatorname*{e}\nolimits^{-i\varphi}$.
In what follows we discuss two regimes of the classical amplification field:
the weak ($\left|  \lambda_{a}\right|  ,\left|  \lambda_{b}\right|  <\left|
\Omega\right|  \ll\Delta$) and the strong ($\left|  \Omega\right|  \gg
\Delta,\left|  \lambda_{a}\right|  ,\left|  \lambda_{b}\right|  $)
amplification regimes. In both cases, the Hamiltonian (\ref{Eq2}) consists of
highly oscillating terms and to a good approximation we finally obtain the
Hamiltonian $\mathcal{H}_{eff}(t)=-i\mathcal{H}(t)\int\mathcal{H}(\tau)d\tau$
\cite{James}, given by%

\begin{align}
\mathcal{H}_{eff}(t)  &  \simeq\frac{1}{\Delta^{2}-\left|  \Omega\right|
^{2}}\left[  \Delta\left|  \lambda_{a}\right|  ^{2}aa^{\dagger}+\Delta\left|
\lambda_{b}\right|  ^{2}b^{\dagger}b-\left|  \Omega\right|  \left(
\lambda_{a}\lambda_{b}\operatorname*{e}\nolimits^{-2i\delta t}ab+\mathrm{h.c.}%
\right)  \right]  \sigma_{ii}\nonumber\\
&  -\frac{1}{2\left(  \Delta-\left|  \Omega\right|  \right)  }\left[  \left|
\lambda_{a}\right|  ^{2}a^{\dagger}a+\left|  \lambda_{b}\right|
^{2}bb^{\dagger}+\left(  \lambda_{a}\lambda_{b}\operatorname*{e}%
\nolimits^{-2i\delta t}ab+\mathrm{h.c.}\right)  \right]  \sigma_{++}%
\nonumber\\
&  -\frac{1}{2\left(  \Delta+\left|  \Omega\right|  \right)  }\left[  \left|
\lambda_{a}\right|  ^{2}a^{\dagger}a+\left|  \lambda_{b}\right|
^{2}bb^{\dagger}-\left(  \lambda_{a}\lambda_{b}\operatorname*{e}%
\nolimits^{-2i\delta t}ab+\mathrm{h.c.}\right)  \right]  \sigma_{--}.
\label{Eq3}%
\end{align}
\textit{PDC in the weak-amplification regime}. To implement this regime it is
sufficient to assume that $\left|  \lambda_{a}\right|  \sim\left|  \lambda
_{b}\right|  $, $\Delta\gtrsim10\times\left|  \lambda_{a}\right|  $ and
$\left|  \Omega\right|  \sim2\left|  \lambda_{a}\right|  $. Starting with the
atomic state prepared in subspace $\left\{  \left|  i\right\rangle \right\}  $
or $\left\{  \left|  \pm\right\rangle \right\}  $, and adjusting $\delta$ such
that $\delta_{i}=-\left(  \left|  \lambda_{a}\right|  ^{2}+\left|  \lambda
_{b}\right|  ^{2}\right)  /\Delta$ or $\delta_{\pm}=\left(  \left|
\lambda_{a}\right|  ^{2}+\left|  \lambda_{b}\right|  ^{2}\right)  /2\left(
\Delta\mp\left|  \Omega\right|  \right)  \approx\left(  \left|  \lambda
_{a}\right|  ^{2}+\left|  \lambda_{b}\right|  ^{2}\right)  /2\Delta$, we
obtain from Eq. (\ref{Eq3}), after the unitary transformation $\exp\left\{
-it\left(  \left|  \lambda_{a}\right|  ^{2}a^{\dagger}a+\left|  \lambda
_{b}\right|  ^{2}b^{\dagger}b\right)  \left[  \sigma_{ii}/\Delta-\sigma
_{++}/2\left(  \Delta-\left|  \Omega\right|  \right)  -\sigma_{--}/2\left(
\Delta+\left|  \Omega\right|  \right)  \right]  \right\}  $, respectively
\begin{subequations}
\label{0}%
\begin{align}
\mathcal{H}_{i}  &  \simeq\left(  \xi_{i}ab+\xi_{i}^{\ast}a^{\dagger
}b^{\dagger}\right)  \sigma_{ii},\label{Eq4a}\\
\mathcal{H}_{\pm}  &  \simeq\left(  \xi_{\pm}ab+\xi_{\pm}^{\ast}a^{\dagger
}b^{\dagger}\right)  \left(  \sigma_{--}-\sigma_{++}\right)  , \label{Eq4b}%
\end{align}
where $\xi_{i}=\Omega^{\ast}\lambda_{a}\lambda_{b}/\Delta^{2}$ and $\xi_{\pm
}=\lambda_{a}\lambda_{b}\operatorname*{e}\nolimits^{i\varphi}/2\left(
\Delta\mp\left|  \Omega\right|  \right)  \approx\lambda_{a}\lambda
_{b}\operatorname*{e}\nolimits^{i\varphi}/2\Delta\equiv\xi$. We observe that
Hamiltonian (\ref{Eq4a}) was used in Ref. \cite{Roberto1} to generate squeezed
states and the original Einstein-Podoslky-Rosen (EPR) entanglement, expanded
in the Fock representation, in two-mode cavity QED{. }However, with the
present technique, Hamiltonian (\ref{Eq4b}) has the advantage that the
coupling parameter $\xi_{\pm}$ is at least one order of magnitude larger than
$\xi_{i}$; consequently, the atom-field interaction time required to obtain a
high-``quality'' EPR state can be considerably shorter,\ making the
dissipative effects negligible. Starting with the atom in the state $\left|
\pm\right\rangle $, both cavity modes being in their vacuum states, and
applying the interaction in Eq. (\ref{Eq4b}) during the time interval $\tau$,
the evolved two-mode state reads
\end{subequations}
\begin{equation}
\operatorname*{e}\nolimits^{-i\mathcal{H}_{\pm}\tau}\left|  \pm\right\rangle
\left|  0,0\right\rangle _{ab}=\left|  \pm\right\rangle \sum_{n=0}^{\infty
}\frac{\left[  \pm\tanh\left(  \left|  \xi\right|  \tau\right)  \right]  ^{n}%
}{\cosh\left(  \left|  \xi\right|  \tau\right)  }\left|  n,n\right\rangle
_{ab}=\left|  \pm\right\rangle \left|  \psi_{\pm}(\tau)\right\rangle
_{ab}\mathrm{{,}} \label{Eq5}%
\end{equation}
where we have adjusted the coupling constants $\lambda_{a}$, $\lambda_{b}$ and
classical phase $\varphi$ so that $\xi=i\left|  \xi\right|  $. State $\left|
\psi_{+}(\tau)\right\rangle _{ab}$ is the two-mode squeezed vacuum state
which, in the limit $\left|  \xi\right|  \tau\rightarrow\infty$ (and projected
into the positional basis of modes $a$ and $b$), is exactly the original
entanglement used in the EPR argument against the uncertainty principle
\cite{EPR}.

To estimate the ``quality'' of the prepared EPR state $\left|  \psi_{+}%
(\tau)\right\rangle _{ab}$ we compute the mean values \cite{SB} $\left(
\Delta x\right)  ^{2}=\left\langle \left(  x_{a}-x_{b}\right)  ^{2}%
\right\rangle =\operatorname*{e}\nolimits^{-2\left|  \xi\right|  \tau}/2$ and
$(\Delta p)^{2}=\left\langle \left(  p_{a}+p_{b}\right)  ^{2}\right\rangle
=\operatorname*{e}\nolimits^{-2\left|  \xi\right|  \tau}/2$, where $x_{\beta
}=\left(  \beta+\beta^{\dagger}\right)  /2$ and $p_{\beta}=-i(\beta
-\beta^{\dagger})/2$ ($\beta=a,b$) are the field quadratures. We obtain\ the
result $\left(  \Delta x\right)  ^{2}+(\Delta p)^{2}=\operatorname*{e}%
\nolimits^{-2\left|  \xi\right|  \tau}$ which goes to zero for the ideal EPR
state ($\left|  \xi\right|  \tau\rightarrow\infty$) and to unity for an
entirely separable state \cite{SB}. Therefore, the expression
$1-\operatorname*{e}\nolimits^{-2\left|  \xi\right|  \tau}$ can be used to
estimate the quality of the prepared EPR state. Assuming typical values for
the parameters involved in cavity QED experiments, we get $\left|  \lambda
_{a}\right|  \sim\left|  \lambda_{b}\right|  \sim3\times10^{5}$s$^{-1}$
\cite{BRH}\textbf{. }With the detuning $\Delta\sim10\times\left|  \lambda
_{a}\right|  $, it follows that $\left|  \xi\right|  \sim1.5\times10^{4}%
$s$^{-1}$, and assuming the interaction time $\tau\sim5\times10^{-5}$s, we
obtain $\left|  \xi\right|  \tau=0.75$ which is larger than the value
$0.69$\ achieved for building the EPR state for unconditional quantum
teleportation in the running-wave domain \cite{Furusawa}. The resulting
quality of the prepared EPR state is $1-\operatorname*{e}\nolimits^{-2\left|
\xi\right|  \tau}$ $\sim0.78$. (This should be compared with the EPR state
engineered through Hamiltonian (\ref{Eq4a}) in Ref. \cite{Roberto1}, where the
above cavity QED parameters lead to $\left|  \xi_{i}\right|  \sim6\times
10^{3}$ and the quality $1-\operatorname*{e}\nolimits^{-2\left|  \xi
_{i}\right|  \tau}$ $\sim0.45$.) The interaction time considered here is about
three (one) orders of magnitude smaller than the field (atom) decay time in
experiments employing closed microwave cavities \cite{Walther}, and about two
(three) orders of magnitude smaller than the field (atom) decay time in
experiments with open microwave cavities \cite{BRH}. Consequently, the
dissipative mechanism becomes negligible. We also note that the proposed
scheme, where classical fields are necessary to induce a Raman transition (see
\cite{Forbidden}), is better suited for open cavities, although closed
microwave cavities can be used as well (see discussion in \cite{Solano}).

Besides having the advantage that $\xi$ $\sim5$ $\xi_{i}$, this technique
makes it is possible to generate mesoscopic superposition states, when
preparing the atom, for example, in the excited state $\left|  e\right\rangle
=\operatorname*{e}\nolimits^{i\varphi/2}\left(  \left|  +\right\rangle
+\left|  -\right\rangle \right)  /\sqrt{2}$. $i)$ In the nondegenerate PDC,
starting with both cavity modes in their vacuum states and applying the
interaction in Eq. (\ref{Eq4b}) during the time interval $\tau$, we generate
the superposition of two-mode vacuum squeezed states $\left|  \psi
(\tau)\right\rangle _{ab}=\mathcal{N}_{\pm}\left(  \operatorname*{e}%
\nolimits^{-i\left|  \Omega\right|  \tau}\left|  \psi_{+}(\tau)\right\rangle
_{ab}\left|  +\right\rangle +\operatorname*{e}\nolimits^{i\left|
\Omega\right|  \tau}\left|  \psi_{-}(\tau)\right\rangle _{ab}\left|
-\right\rangle \right)  $. Adjusting $\left|  \Omega\right|  \tau=2\pi$ and
measuring the atomic state after the atom-field interaction, we obtain the
even and odd EPR states, which we define as
\begin{equation}
\left|  \Psi_{\binom{even}{odd}}\right\rangle _{ab}=\mathcal{N}_{\pm}\left(
\left|  \psi_{+}(\tau)\right\rangle _{ab}\pm\left|  \psi_{-}(\tau
)\right\rangle \right)  =\mathcal{N}_{\pm}\sum_{n=0}^{\infty}\left[
1\pm\left(  -1\right)  ^{n}\right]  \frac{\left[  \tanh\left(  \left|
\xi\right|  \tau\right)  \right]  ^{n}}{\cosh\left(  \left|  \xi\right|
\tau\right)  }\left|  n,n\right\rangle _{ab}{.} \label{Eq6}%
\end{equation}
Similarly to the even and odd coherent states \cite{YS}, $\left\langle
\psi_{\pm}(\tau)\right.  \left|  \psi_{\mp}(\tau)\right\rangle =\cosh
^{-1}(2\left|  \xi\right|  \tau)\sim2\operatorname*{e}\nolimits^{-2\left|
\xi\right|  \tau}$ for $2\left|  \xi\right|  \tau\gg1$, while $\left\langle
\Psi_{even}\right.  \left|  \Psi_{odd}\right\rangle _{ab}$ $=0$. $\ ii)$ In
the degenerate PDC ($\omega_{a}=\omega_{b}$), the engineered cavity-field
superposition, after the atom-field interaction and the measurement of the
atomic state, is written as $|\Phi\left(  \tau\right)  \rangle=\mathcal{N}%
_{\pm}\left[  \operatorname*{e}^{-i\left|  \Omega\right|  \tau}S(\xi,\tau
)\pm\operatorname*{e}^{i\Omega\tau}S^{-1}(\xi,\tau)\right]  |\Phi\left(
0\right)  \rangle$, where $S(\xi,\tau)=\exp\left[  -i\left(  \xi a^{\dagger
2}+\xi^{\ast}a^{2}\right)  \tau\right]  $ stands for the squeeze operator
(with the squeezing factor $r=2\left|  \xi\right|  \tau$), and the $+$ ($-$)
sign occurs if the atom is detected in state $|e\rangle$ ($|g\rangle$). We
note that the components of the superposition $|\Phi\left(  \tau\right)
\rangle$ are squeezed in perpendicular directions. Besides the large coupling
strength $\left|  \xi\right|  $ achieved, another advantage of the present
scheme to generate the superpositions in Eq. (\ref{Eq6}) and $|\Phi\left(
\tau\right)  \rangle$ is that no Ramsey zones or external parametric
amplification are required. We also note that the degenerate PDC can be
implemented considering non-circular Rydberg states in an appropriated
configuration as discussed in Ref. \cite{Brune}.

To estimate the degree of squeezing achieved with the present scheme we
consider the degenerate PDC, starting with the cavity mode prepared in a
coherent state $\left|  \alpha\right\rangle $ and the atom prepared in the
state $\left|  +\right\rangle $. The variance in the squeezed quadrature of
the generated coherent squeezed state $S(\xi,\tau)\left|  \alpha\right\rangle
$ is $\left\langle \Delta X\right\rangle ^{2}=\operatorname*{e}^{-4\left|
\xi\right|  \tau}/4$. Assuming the typical cavity-QED values defined above for
$\left|  \lambda_{a}\right|  $, $\left|  \lambda_{b}\right|  $, $\Delta$, and
$\tau$, we finally obtain the squeezing factor $r=2\left|  \xi\right|  \tau$
$=1.5$. With such values, the variance in the squeezed quadrature turns out to
be $\left\langle \Delta X\right\rangle ^{2}\sim1.2\times10^{-2}$, representing
a squeezing around $95\%$.

\textit{PDC in the strong-amplification regime}. Here we assume that $\left|
\lambda_{a}\right|  \sim\left|  \lambda_{b}\right|  \gtrsim\Delta$ and
$\left|  \Omega\right|  $ $\gtrsim10\times\left|  \lambda_{a}\right|  $. For
the atomic state prepared in subspace $\left\{  \left|  i\right\rangle
\right\}  $ or $\left\{  \left|  \pm\right\rangle \right\}  $, and assuming
$\delta=0$, we obtain from (\ref{Eq3}) the Hamiltonians
\begin{subequations}
\label{0}%
\begin{align}
\mathcal{H}_{i}  &  \simeq-\left(  \zeta_{i}ab+\zeta_{i}^{\ast}a^{\dagger
}b^{\dagger}\right)  \sigma_{ii},\label{Eq7a}\\
\mathcal{H}_{\pm}  &  \simeq\frac{1}{2\left|  \Omega\right|  }\left(  \left|
\lambda_{a}\right|  ^{2}a^{\dagger}a+\left|  \lambda_{b}\right|
^{2}bb^{\dagger}\right)  \left(  \sigma_{++}-\sigma_{--}\right)  +\left(
\zeta_{\pm}ab+\zeta_{\pm}^{\ast}a^{\dagger}b^{\dagger}\right)  \left(
\sigma_{++}+\sigma_{--}\right)  , \label{Eq7b}%
\end{align}
where $\zeta_{i}=\lambda_{a}\lambda_{b}\operatorname*{e}\nolimits^{i\varphi
}/\left|  \Omega\right|  $ and $\zeta_{\pm}=\lambda_{a}\lambda_{b}/2\left(
\left|  \Omega\right|  \mp\Delta\right)  \approx\lambda_{a}\lambda
_{b}/2\left|  \Omega\right|  \equiv\zeta$. Differently from the
weak-amplification regime, here we may assume $\Delta=0$. Note that apart from
the global phase, Hamiltonian (\ref{Eq7a}) is exactly the same as
(\ref{Eq4a}), but with the advantage that $\zeta_{i}\sim10\times\xi_{i}$. With
the cavity-QED experimental parameters defined above, we obtain $\left|
\zeta_{i}\right|  \tau=1.5$, so that the EPR state generated through
Hamiltonian (\ref{Eq7a}) has the quality{\LARGE \ }$1-\operatorname*{e}%
\nolimits^{-2\left|  \zeta_{i}\right|  \tau}${\LARGE \ }$\sim0.95$. Hence, in
the degenerate PDC, the squeezing factor obtained when engineering a squeezed
coherent state from Hamiltonian (\ref{Eq7a}) is $2\left|  \zeta_{i}\right|
\tau=3.0$ and the variance in the squeezed quadrature becomes $\left\langle
\Delta X\right\rangle ^{2}\sim6.2\times10^{-4}$, leading to a squeezing of
$99.7\%$.

Hamiltonian (\ref{Eq7b}) also possesses considerable advantages, compared to
Eq. (\ref{Eq4a}), even though $\zeta\sim\xi$ (for the parameter values assumed
so far). Considering again the degenerate PDC process and the atom prepared in
the excited state $\left|  e\right\rangle =\operatorname*{e}%
\nolimits^{i\varphi/2}\left(  \left|  +\right\rangle +\left|  -\right\rangle
\right)  /\sqrt{2}$, the resulting Hamiltonian, leading to two uncoupled
evolutions for the atom-field system, depending on the atomic state $\left|
+\right\rangle $ or $\left|  -\right\rangle $, is written in the
Schr\"{o}dinger picture as $\mathcal{H}_{\pm}=\left(  \omega\pm\chi\right)
a^{\dagger}a+\left(  \zeta\operatorname*{e}\nolimits^{-2i\omega t}\left(
a^{\dagger}\right)  ^{2}+\zeta^{\ast}\operatorname*{e}\nolimits^{2i\omega
t}a^{2}\right)  $, with $\chi=\left(  \left|  \lambda_{a}\right|  ^{2}+\left|
\lambda_{b}\right|  ^{2}\right)  /2\left|  \Omega\right|  $ ($\sim2\left|
\zeta\right|  $). This Hamiltonian was analyzed in detail in Ref. \cite{CFR1},
using the time-dependent invariants by Lewis and Riesenfeld \cite{LR} and for
the values presented above it corresponds to the case of critical coupling,
where $\left|  \zeta\right|  /\chi=1$. Starting from the initial state
$\operatorname*{e}\nolimits^{i\varphi/2}\left(  \left|  +\right\rangle
+\left|  -\right\rangle \right)  \left|  \alpha\right\rangle /\sqrt{2}$,
$\left|  \alpha\right\rangle $ being a coherent state injected into the
cavity, the evolved superposition reads $\left(  \left|  +\right\rangle
U_{+}+\left|  -\right\rangle U_{-}\right)  \left|  \alpha\right\rangle
/\sqrt{2}$, where $U_{\pm}$ stands for the evolution operator associated with
Hamiltonian $\mathcal{H}_{\pm}$, as derived in Ref. \cite{CFR1}. After
interacting with the cavity mode, during the time interval $\tau$, the atomic
state is measured and we finally obtain the cavity-field superposition
\end{subequations}
\begin{equation}
\left|  \Psi(\tau)\right\rangle =\mathcal{N}_{\pm}\left(  \operatorname*{e}%
\nolimits^{-i\left|  \Omega\right|  \tau}U_{+}\pm\operatorname*{e}%
\nolimits^{i\left|  \Omega\right|  \tau}U_{-}\right)  \left|  \alpha
\right\rangle , \label{Eq8}%
\end{equation}
where the sign $+$($-$) arises from the detection of the state $\left|
e\right\rangle $ ($\left|  g\right\rangle $). The interesting feature of this
scheme for engineering a squeezed ``Schr\"{o}dinger cat''-like state\ is that
no Ramsey zones or external parametric amplification field are needed,
differently from Ref. \cite{CFR1}. The squeezing factor achieved,
$r=\operatorname{arcsinh}(2\left|  \zeta\right|  \tau)$, computed in Ref.
\cite{CFR1}, is around $1.2$, assuming typical parameters in cavity QED. It
has been shown in Ref.\cite{CFR1} that the decoherence time of such mesoscopic
superposition could be around the relaxation time of the cavity field when
assuming a particular squeezed reservoir at absolute zero. This reservoir must
be composed of oscillators squeezed in a direction perpendicular to that of
the superposition state. Therefore, in this paper we have solved partially the
problem of engineering a truly mesoscopic state in cavity QED. The task of
engineering an optimal squeezed reservoir in cavity QED remains to be
achieved. In Ref. \cite{Vitali} the authors show how to prepared a
partially-squeezed reservoir in cavity QED.

\textit{PUC}. Now, the energy diagram of the Rydberg three-level atom is in
the $\Lambda$ configuration, where the ground and excited states are coupled
through an auxiliary more-excited level, as in Fig. 1b. The cavity microwave
modes of frequencies $\omega_{a}$ and $\omega_{b}$ ( $\omega_{a}-\omega
_{b}=2\omega_{0}$) enable both dipole-allowed transitions $\left|
g\right\rangle $ $\leftrightarrow$ $\left|  i\right\rangle $ and $\left|
e\right\rangle $ $\leftrightarrow$ $\left|  i\right\rangle $, with coupling
constants $\lambda_{a}$ and $\lambda_{b}$, respectively, and detuning $\Delta$
$=\omega_{i}+\omega_{0}-\omega_{a}=\omega_{i}-\omega_{0}-\omega_{b}$.
(Evidently, in this case we can assume at most two levels as circular Rydberg
states.) A classical field of frequency $\omega=2\left(  \omega_{0}%
-\delta\right)  $, driving the atomic transition $\left|  g\right\rangle $
$\leftrightarrow$ $\left|  e\right\rangle $ with coupling constant $\Omega$,
leads to the desired interaction between the modes $\omega_{a}$ and
$\omega_{b}$. Within the rotating wave approximation, the Hamiltonian
$H=H_{0}+V(t)$, is given by
\begin{subequations}
\label{Eq9}%
\begin{align}
H_{0}  &  =\hbar\omega_{a}a^{\dagger}a+\hbar\omega_{b}b^{\dagger}b+\hbar
\omega_{g}\sigma_{gg}+\hbar\omega_{e}\sigma_{ee}+\hbar\omega_{i}\sigma
_{ii},\label{Eq9a}\\
V(t)  &  =\hbar\left(  \lambda_{a}a\sigma_{ig}+\lambda_{b}b\sigma_{ie}+\Omega
e^{-i\omega t}\sigma_{ge}+\mathrm{{h.c.}}\right)  \mathrm{{.}} \label{Eq9b}%
\end{align}
Following analogous steps to those used in the above analysis for the PDC, we
obtain the effective Hamiltonian for the PUC.

\textit{PUC in the weak-amplification regime}. From an atomic state prepared
in subspace $\left\{  \left|  i\right\rangle \right\}  $ or $\left\{  \left|
\pm\right\rangle \right\}  $, adjusting $\delta$ as $\delta_{i}=\left(
\left|  \lambda_{b}\right|  ^{2}-\left|  \lambda_{a}\right|  ^{2}\right)
/2\Delta$ or $\delta_{\pm}=\left(  \left|  \lambda_{a}\right|  ^{2}-\left|
\lambda_{b}\right|  ^{2}\right)  /4\left(  \Delta\mp\left|  \Omega\right|
\right)  \approx\left(  \left|  \lambda_{a}\right|  ^{2}-\left|  \lambda
_{b}\right|  ^{2}\right)  /4\Delta$, we obtain, respectively
\end{subequations}
\begin{subequations}
\label{Eq10}%
\begin{align}
\mathcal{H}_{i}  &  \simeq\left(  \gamma_{i}ab^{\dagger}+\gamma_{i}^{\ast
}a^{\dagger}b\right)  \left|  i\right\rangle \left\langle i\right|
,\label{Eq10a}\\
\mathcal{H}_{\pm}  &  \simeq\left(  \gamma_{\pm}ab^{\dagger}+\gamma_{\pm
}^{\ast}a^{\dagger}b\right)  \left(  \left|  -\right\rangle \left\langle
-\right|  -\left|  +\right\rangle \left\langle +\right|  \right)  ,
\label{Eq10b}%
\end{align}
where $\gamma_{i}=\left|  \Omega\right|  \lambda_{a}\lambda_{b}^{\ast
}\operatorname*{e}\nolimits^{i\varphi}/\Delta^{2}$ and $\gamma_{\pm}%
=\lambda_{a}\lambda_{b}^{\ast}\operatorname*{e}\nolimits^{i\varphi}/2\left(
\Delta\mp\left|  \Omega\right|  \right)  \approx\lambda_{a}\lambda_{b}^{\ast
}\operatorname*{e}\nolimits^{i\varphi}/2\Delta\equiv\gamma$. After preparing
the atom in the ground state and mode $\omega_{a}$ ($\omega_{b}$) in the
coherent state $\left|  \alpha\right\rangle _{a}$ ($\left|  \beta\right\rangle
_{b}$), the interaction (\ref{Eq10b}) can be used to generate cavity-field
entangled states (without the need for Ramsey zones) such as $\left|
\psi(\tau)\right\rangle _{ab}=\left(  \operatorname*{e}\nolimits^{\left|
\gamma\right|  \tau\left(  ab^{\dagger}-a^{\dagger}b\right)  }\left|
+\right\rangle +\operatorname*{e}\nolimits^{-\left|  \gamma\right|
\tau\left(  ab^{\dagger}-a^{\dagger}b\right)  }\left|  -\right\rangle \right)
\left|  \alpha,\beta\right\rangle _{ab}/\sqrt{2}$, where we have assumed
$\gamma=-i\left|  \gamma\right|  $. Adjusting the atom-field interaction time
so that $\left|  \gamma\right|  \tau=\pi/2$, we obtain, after the atomic-state
detection, $\left|  \psi_{\pm}(\tau)\right\rangle _{ab}=\mathcal{N}_{\pm
}\left(  \operatorname*{e}\nolimits^{-i\left|  \Omega\right|  \tau}\left|
\beta,-\alpha\right\rangle \pm\operatorname*{e}\nolimits^{i\left|
\Omega\right|  \tau}\left|  -\beta,\alpha\right\rangle \right)  $, where the
$+$ ($-$) sign occurs if the atom is detected in state $|e\rangle$
($|g\rangle$).

\textit{PUC in the strong-amplification regime}. In this case, we obtain
exactly the Hamiltonians in Eqs. (\ref{Eq10a}) and (\ref{Eq10b}), but with
coupling strengths $\eta_{i}=\lambda_{a}\lambda_{b}^{\ast}\operatorname*{e}%
\nolimits^{i\varphi}/\left|  \Omega\right|  $ and $\eta_{\pm}=\mp\lambda
_{a}\lambda_{b}^{\ast}\operatorname*{e}\nolimits^{i\varphi}/2\left|
\Omega\right|  $. Since $\eta_{i}$ is around an order of magnitude larger than
$\gamma_{i}$, beam-splitter operations can be realized in the
strong-amplification regime with a smaller atom-field interaction time.

Finally, we discuss some possible error sources in our scheme, starting with
the validaty of the effective Hamiltonians coming from the approximation
$\mathcal{H}_{eff}(t)=-i\mathcal{H}(t)\int\mathcal{H}(\tau)d\tau$. In order to
compare the evolutions governed by the effective Hamiltonians $\mathcal{H}%
_{i}$ and $\mathcal{H}_{\pm}$, for the PUC, with those calculated without this
approximation, defined in Eqs (\ref{Eq9a}) and (\ref{Eq9b}), we computed
numerically the time evolution of the initial states $\left|  i\right\rangle
\left|  1\right\rangle _{a}\left|  0\right\rangle _{b}$ and $\left|
\pm\right\rangle \left|  1\right\rangle _{a}\left|  0\right\rangle _{b}$ in
both cases. For the experimental parameters defined above, the divergence
between the curves is about $5\%$ in both regimes of amplification. Evidently,
when considering a ratio $\Delta/\left|  \lambda_{a}\right|  >10$ ($\left|
\Omega\right|  /\left|  \lambda_{a}\right|  >10$) in the weak (strong)
coupling regime, the approximation becomes even better, at the expense of a
larger atom-field interaction time.

Focusing on the squeezed coherent state engineered in the degenerate PDC in
the strong-amplification regime, we have estimated the squeezing factor
$\widetilde{r}=2\left\vert \zeta_{i}\right\vert (1-\operatorname*{e}%
\nolimits^{-\Gamma_{a}\tau})/\Gamma_{a}$ and the variance of the squeezed
quadrature $\left\langle \Delta\widetilde{X}\right\rangle ^{2}=\left[
1-\left(  1-\operatorname*{e}\nolimits^{-2\widetilde{r}}\right)
\operatorname*{e}\nolimits^{-\Gamma_{c}\tau}\right]  /4$ taking into account
the cavity damping rate ($\Gamma_{c}$) and the atomic decay ($\Gamma_{a}$),
which are introduced phenomenologically following the reasoning in Ref.
\cite{Celso1}. Considering non-circular Rydberg states in experiments with
open cavities, such that $\Gamma_{c}\sim10^{3}$\ s$^{-1}$ \cite{BRH} and
$\Gamma_{a}\sim5\times10^{3}$ s$^{-1}$, together with the values assumed above
and an atom-field interaction time about $\tau\sim5\times10^{-5}$s, we obtain
$\widetilde{r}\sim2.6$ and $\left\langle \Delta X\right\rangle ^{2}%
\sim1.3\times10^{-2}$, representing a squeezing around $95\%$. As mentioned
above, for the atom-field interaction time required in our technique, the
dissipative mechanism becomes practically negligible. We finally mention that
a sample of $N$ identical atoms may be employed to enhance the squeezing
factor as recently considered by Guzman et al. \cite{Guzman}.

\end{subequations}
\begin{acknowledgments}
We wish to express thanks for the support from FAPESP and CNPq (Instituto do
Mil\^{e}nio de Informa\c{c}\~{a}o Qu\^{a}ntica), Brazilian agencies. We also
thank Profs. L. Davidovich, N. Zagury, R. L. de Matos Filho, N. G. de Almeida,
R. M. Serra, and P. A. Nussenzveig.
\end{acknowledgments}

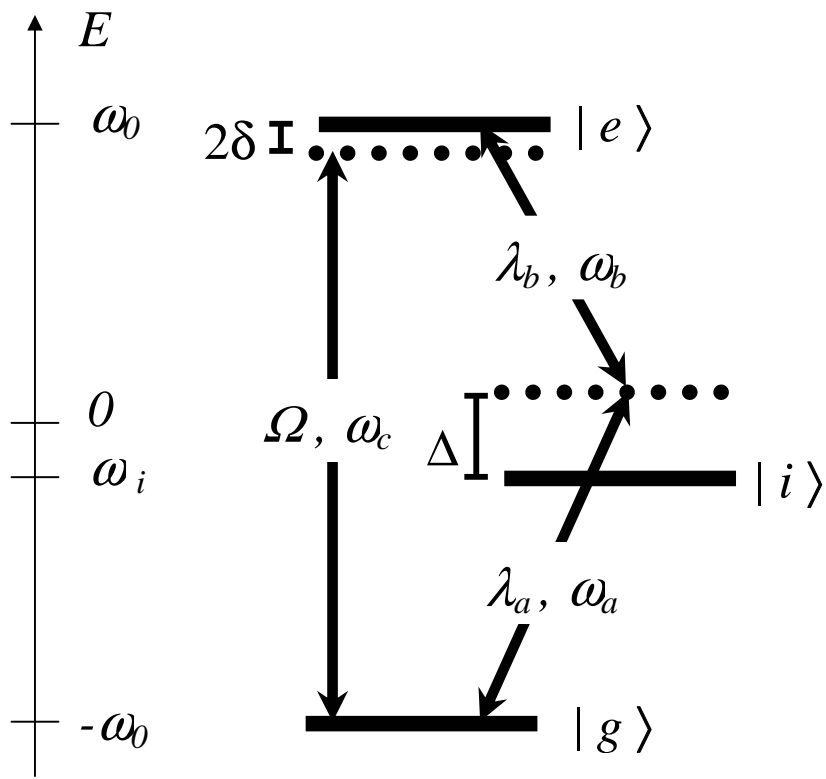
Fig. 1. Energy diagram of a three-level atomic system in the (a) ladder and
(b) lambda configurations.1


\begin{thebibliography}{99}
\bibitem {Forbidden}This dipole-forbidden transition can be induced by taking
advantage of dipole-allowed transitions to another auxiliary level. In this
case, a fourth level $\left\vert f\right\rangle $ must be added to the atomic
system together with appropriated classical fields inducing a Raman transition
with effective coupling given by $\Omega=g_{1}^{\ast}g_{2}/\widetilde{\Delta}%
$, $g_{1}$ and $g_{2}$ being the coupling constants to the vicinity of the
dipole-allowed transitions $\left\vert g\right\rangle $ $\leftrightarrow$
$\left\vert f\right\rangle $ and $\left\vert e\right\rangle $ $\leftrightarrow
$ $\left\vert f\right\rangle $, both with detuning $\widetilde{\Delta}\gg$
$\left\vert g_{1}\right\vert $,$\left\vert g_{2}\right\vert $. For the strong
coupling regime it is required that $\Omega\sim10^{6}$s$^{-1}$, which follows
from $g_{1}\sim g_{2}\sim10^{7}$s$^{-1}$ (easily achieved for dippole-allowed
transitions). Note that we may also consider the auxiliary level $\left\vert
i\right\rangle $ itself for this purpose, watching out for the classical
fields to be far from resonance with the cavity modes.
\end{thebibliography}
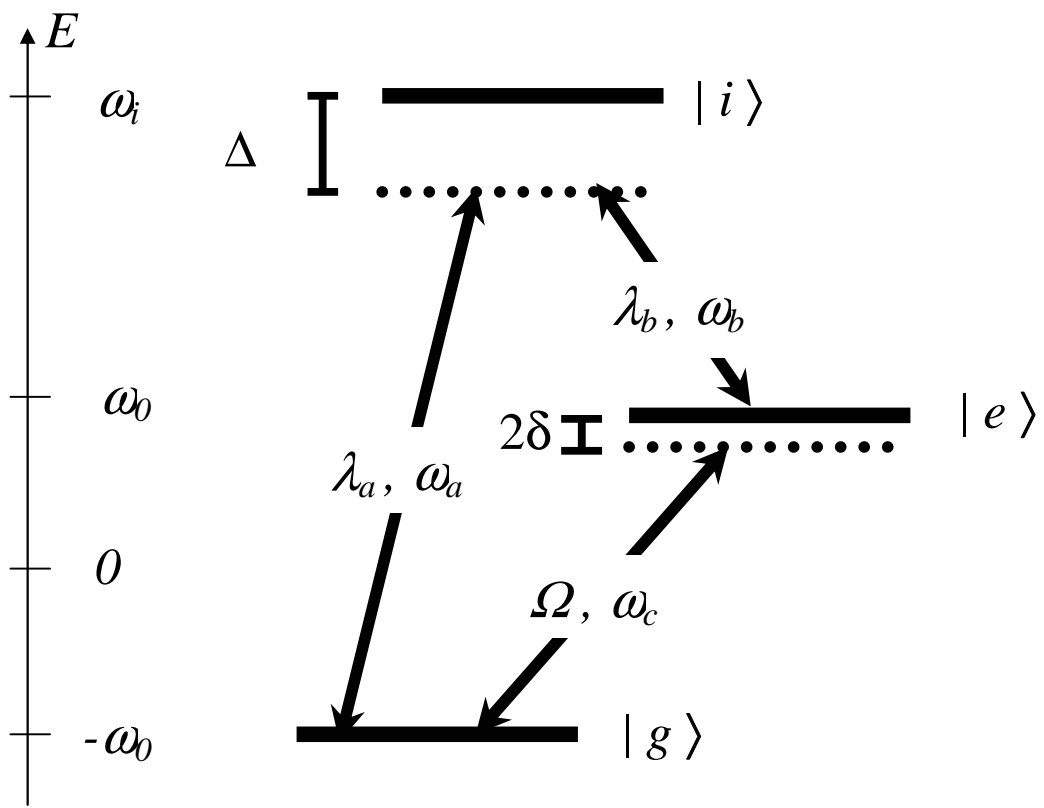
\end{document}